\documentstyle[11pt]{article}

\begin{document}

{\scriptsize \em
3rd.April 2002}

\noindent
J. Superconductivity: Incorporating Novel Magnetism, to be published

\begin{center}
{\large{\bf Interpretation of large room-temperature diamagnetism at
low magnetic fields in films of oxidised atactic polypropylene in terms
of superconducting current loops }}

\vskip 2em
{\bf D.M. Eagles$^{1}$}

\vskip 2em
\end{center}

\vskip 2em
\noindent
{\bf Abstract}

\begin{quote}

A simple model is used to analyse published results on large
room-temperature diamagnetism for two films of oxidised atactic
polypropylene (OAPP) at low magnetic fields.  The model involves induced
currents expected in circular closed loops of superconductors in fields
below the lower critical field $H_{c1}$ at which flux penetration would
first occur if a metamagnetic transition did not intervene as in OAPP,
and the assumption that resistance would be restored at $H_{c1}$
(negligible pinning).  Fits to the data for the more strongly magnetic
sample with the model, allowing two different types of loops with
different loop radii $b_1$ and $b_2$, but with the same cross section
$a$ of loop material yield $H_{c1} \approx 5260$ Oe, and fits to the
data for the less strongly magnetic sample with two loop sizes and with
the same value of $H_{c1}$, combined with the knowledge that the
minimum number of closed loops of any type is one, requires that the
radius $a$ of the cross section of the material should be less than
about $0.8 \mu$m,  in fair agreement with a maximum radius of 1 $\mu$m
obtained previously from other data.

\end{quote}


\noindent
$^{1}$   
19 Holt Road, Harold Hill, Romford, Essex RM3 8PN, England.
e-mail:  d.eagles@ic.ac.uk
\newline

\noindent
{\bf 1. INTRODUCTION}

There is evidence  from at least three different types of experiments
that narrow channels through films of oxidised atactic polypropylene
(OAPP) are superconducting at room temperature.  The three types of
evidence are (i) lower limits for conductivity several orders of
magnitude greater than that of copper, found by direct [1] and indirect [2]
methods, (ii) non-thermal destruction of ultra-high conductivity by
high pulsed currents, with critical current densities greater than
$10^9$ A cm$^{-2}$ [2], and (iii) negligible electronic contribution to
the thermal conductivity [3].  Further support for high-temperature
superconductivity in channels of a different polymer is that the
thermopower between 87 and 233 K in films of poly(octylmethacrylate) is
zero to within estimated errors of measurements [4].

The magnetic properties of films of OAPP are also unusual. The unusual
properties observed in samples in which highly conducting channels
occur are (i) a metamagnetic transition in fields of a few kilooersted
[5-7], (ii) large diamagnetism observed at low fields in about
10\% of the samples showing a metamagnetic transition [6,7], and (iii)
spontaneous forces occurring in some field range tending to push
samples to lower magnetic field regions in inhomogeneous magnetic
fields [8].

At least three different models have been suggested for the ultra-high
conductivity or the superconductivity [9-14], two for the high
critical current densities in channels [10-14], and two for the unusual
magnetic properties [6,7,11,12].  Authors of papers on both types of
models for the magnetic properties are agreed that the large
diamagnetism in some samples and the occasional spontaneous forces
pushing samples out to low-field regions in inhomogeneous fields are 
associated with superconducting channels which form closed loops, but differ
as to what is happening in these closed loops.   In [11,12] I suggested
that large spontaneous currents occur when closed loops form, and gave
support for this hypothesis on the basis that some data in a figure in
[15] appear to show that the susceptibility is approximately
proportional to (1/field), indicating a constant moment, independent of
field, whereas the authors of [7] appear to think the large
diamagnetism  (of the order of a percent of a complete Meissner effect
at low fields in one sample) is associated with the percentage of the
film which is superconducting.  We think that this hypothesis is not
compatible with the assumption that the diamagnetism is associated with
closed loops, since the total fraction of material occupied by
conducting channels is typically at most a few percent (channels
separated by 7-8 $\mu$m estimated in one very thin film, thickness 0.3
$\mu$m, showing conducting channels [16], and channel diameters always
less than 2 $\mu$m [17,2]), the fraction of material in closed loops
can be expected to be considerably less than this, and also the
Meissner effect may be further reduced if the magnetic-field
penetration depth is comparable to channel radii.  However, recent
experimental results [7] do not appear to support the spontaneous
moment hypothesis well either, as there is at best only a small range
of fields for which the diamagnetic moment is approximately constant.

In this paper we explore the hypothesis that the diamagnetic moment is
associated with induced supercurrents at fields below which the loop
becomes resistant.  We make use of a theory for induced currents in
circular loops of Type I superconductors below the critical field,
discussed in Shoenberg's book (1952 version) [18].  Looking at the
derivation of his results, it appears that they will apply to Type II
superconductors below $H_{c1}$ if there is negligible flux line
pinning, so that resistance is restored as soon as fields reach
$H_{c1}$.  A metamagnetic transition may occur before $H_{c1}$ is
reached, but we assume that the metamagnetic transition, being a
cooperative phenomenon, will not occur until the average field
throughout the material of the loop reaches a certain value $H_M$, and
so supercurrents can continue to flow even if the field near the outer
surface of the loop is greater than $H_M$, provided that the average
field in the material is smaller than this value.  

Since only a small fraction of the conducting channels
form closed loops, we assume that the observed diamagnetism at low
fields is a superposition of a diamagnetism associated with closed
loops, and a smaller positive magnetisation as found by a combination
of results of observations made on the next day after closed loops have
been destroyed by large fields and extrapolation.  Thus we fit the
larger diamagnetism found after correcting for this effect.

\noindent {\bf 2. FITTING DATA WITH TWO TYPES OF
SUPERCONDUCTING LOOPS}

Let us introduce the notation $H_R$ for the field at which resistance
appears in the superconducting channels of which the loops are made,
either $H_{c}$ for a Type I superconductor, or $H_{c1}$ for a Type II
superconductor with no pinning. 
Let us define a field 
\begin{equation}
H_B = 0.5 H_R.  
\end{equation} 
Then Shoenberg's type of theory [18] shows that, for a circular loop of
superconductor of radius $b$ with a circular cross section of material
of radius $a$, initially in zero field, the magnetisation commences to
vary linearly with applied field $H_e$ perpendicular to the plane of
the loop to a value
\begin{equation} 
m_A = -\frac{\pi ab^2 H_B}{1+La/\pi b^2} 
\end{equation} 
when the field reaches a value $H_A$ given by
\begin{equation} 
H_A = \frac{H_B(La/\pi b^2)}{1+La/\pi b^2}, 
\end{equation} 
where $L$ is the inductance of the loop.  The
inductance is given by 
\begin{equation} 
L = 4\pi b X, 
\end{equation}
where, for supercurrents confined to the surface of the material of the
loop, 
\begin{equation} 
X = [\rm{ln}(8b/a)-2], 
\end{equation} 
whereas for supercurrents through the bulk of the material of the loop,
\begin{equation} 
X = [\rm{ln}(8b/a)-7/4].  
\end{equation} 
For, e.g.  $b/a$ = 50, the difference in the two values of $X$ is
6\%.  For definiteness we shall do our calculations with the first value
of $X$, which will be correct if the magnetic-field penetration depth
is small compared with $a$.  The initial susceptibility $\chi$
associated with the loop is
\begin{equation}
\chi = -\pi b^3 /4 X.
\end{equation}

As the applied field is increased above $H_A$, the magnitude of the
magnetic moment decreases linearly with field until the field $H_B$ is
reached, at which field the magnetic moment is
\begin{equation}
m_B = -\pi b a^2 H_B.
\end{equation}
At fields above $H_B$, ignoring a small discontinuity in moment at
$H_B$ mentioned by Shoenberg but not discussed in detail, there is a
linear decrease of the magnitude of the magnetic moment from the value
given by Eq. (7) to zero at a field approximately equal [to lowest
order in $(a/4b)$] to $2 H_B = H_R$.

In order to fit the observed moments starting from zero applied field
and continuing up to the field at which the metamagnetic transition
occurs for the sample of [7] with the largest moment, we use a model
with two types of loops with different $b/a$ ratios,
\begin{equation}
r_1 = b_1/a > b_2/a = r_2,
\end{equation} 
with $a$ the same for both loop types, and suppose that there are $N_1$
and $N_2$ loops of each type.  Such a model is used as a first
approximation to a model with a semicontinuous distribution of loop
parameters.  We also assume that the planes of the loops are parallel
to the film surfaces, i.e. perpendicular to the applied field.  For the
larger loops, geometrical consraints for a film of thickness 10 $\mu$m
[7] force any closed loops to have approximately this orientation.  If
some of the smaller loops have other orientations, then probably the
theory will still hold approximately with the reinterpretation of loop
areas as the average projections of their areas on to planes parallel
to the film surfaces.

The model with two types of loop parameters has four straight-line
segments on the moment versus field curve, with discontinuities in
slope at fields $H_{A1}$, $H_{A2}$, and $H_B$, where $H_{A1}$ and
$H_{A2}$ are given by Eq. (3) for the two different values of $b/a$.
The magnetic moments at the points of discontinuity of slope are given
by
\begin{equation}
m(H_{A1})= N_1 m_{A1} + N_2 m_{A2} (H_{A1}/H_{A2}),
\end{equation}
\begin{equation}
m(H_{A2})= N_1 m_{A1}-N_1(m_{A1}-m_{B1})
(\frac{H_{A2}-H_{A1}}{H_B-H_{A1}}) + N_2 m_{A2},
\end{equation} 
\begin{equation}
m(H_B) = N_1 m_{B1}+N_2 m_{B2},
\end{equation}
with the magnetisation going to zero at $H_R=2H_B$.  In Eqs. (10) to (12),
the second suffices 1 and 2 on $m_A$ and $m_B$ refer to the two types
of loops.   

After correcting for a probable effect of superposition of diamagnetism
associated with closed loops with a smaller positive moment associated
with the majority of conducting channels which do not form closed
loops, as discussed in the introduction, we fit the data for the more
strongly magnetised sample with five adjustable parameters, viz.
$(b_1/a)$, $(b_2/a)$ (assuming $b_1>b_2$), $N_1m_{A1}$, $N_2m_{A2}$,
and $H_B$.  We find that $b_1/a=133$, $b_2/a=7.9$, $H_B=2631$ Oe,
implying $H_{c1}=5262$ Oe, and, with use of the fact that the volume of
the film $10^{-3}$ cm$^{3}$ [7], that $N_1 m_{A1}= -1.69\times 10^{-4}$
emu, and $N_2 m_{A2} = -1.25\times 10^{-4}$ emu.  Using Eqs. (2), (4)
and (5) we deduce that $N_2/N_1 \approx 380$, corresponding to a ratio
of area covered by the smaller loops to that covered by the larger
loops of about 1.3.  We used a program AMOEBA given in a book [19] to
perform the least squares fitting.  The fit to the data up to the
field of the metamagnetic transition ($\approx 3360$ Oe for this
sample) is shown in Fig. 1.  The rms accuracy of the fit is 2.2\% of
the mean value of the inferred diamagnetism, or 3.3\% of the mean value
of the net observed magnetisation.

For the second sample we assume that the corrections for a positive
moment from the channels which do not form closed loops have values
half of those for the first sample, based on the estimate in [7] that
the average electron concentration in the second sample is 
about half that of the first sample.  We use use the same type of
model with two different values of $(b/a)$,  but keep $H_B$ as before,
and so we have a four-parameter fit.  We find $b_1/a=106$,
$b_2/a=14.4$, and, with the volume of the film as $10^{-3}$ cm$^{3}$
[7], $N_1m_{A1}=-0.35\times 10^{-4}$ emu, $N_2m_{A2}=-0.25\times
10^{-4}$ emu.  From Eqs. (2), (4) and (5) we deduce that $N_2/N_1
\approx 58$, corresponding to a ratio of the area covered by the
smaller loops to that covered by the larger loops of about 1.1.  The
rms accuracy of the fit is 2.4\% of the mean value of the inferred
diamagnetism, or 6.3\% of mean value of the net observed
magnetisation.  The fit to the data up to the field of about 1390 Oe
for the metamagnetic transition for this sample is also shown in Fig.
1.  Since $N_1$ cannot be less than 1, we deduce  from Eqs. (2), (4)
and (5) and the value of $N_1 m_{A1}$ that the radius $a$ of the cross
sections of the channels forming the loops is less than 0.76 $\mu$m.
This is in fair agreement with an upper limit of 1 $\mu$m estimated
from other data [17,2].  A lower limit for channel radii of 0.1 $\mu$m
is estimated in [20], and a stricter upper limit of 0.35 $\mu$m is
mentioned in [2], but with only a reference to a future publication
(which has not appeared yet as far as I know) for an explanation of how
this limit is obtained.

From the parameters obtained we find that the fraction of the first
sample occupied by the material of the closed loops is $7.9 \times
10^{-5}$ for the set of smaller loops and $0.35 \times 10^{-5}$ for the
set of larger loops, independent of the value of $a$.  The total is
sufficiently small compared with the probable total fractional volume
of the order of a percent occupied by all conducting channels that we
are justified in using approximately the same correction for positive
magnetisation as that for all channels after the closed loops are
broken, inferred from magnetisation measurements on the day after the
original measurements [7].  The contributions to the initial
susceptibility from the larger and smaller loops are $-4.9 \times
10^{-4}$ emu cm$^{-3}$ and $-0.9 \times 10^{-4}$ emu cm$^{-3}$
respectively, corresponding to a total of 0.7\% of that for a complete
Meissner effect.   The appreciably smaller fraction of superconducting
material compared with the fraction of the full Meissner effect for the
initial susceptibility arises because the susceptibility due to induced
currents for a closed loop is larger by a factor of $(1/2X)(b/a)^2$
than that due to a Meissner effect keeping the flux completely out of
the material of the loop.  Although, for the model used, most of the
material of the superconducting loops is associated with the smaller
loops, the dominant contribution to the initial susceptibility comes
from the larger loops because of the factor $(b/a)^2$.

For $a = 0.76\mu$m, the maximum induced currents in the loops, at the
fields $H_{Ai}$ (i=1,2), vary between 1.0 A and 1.7 A, depending on the
loop size.  These currents are considerably smaller than the critical
currents of about 60 A through films in pulsed measurements with
microprobe contacts of diameter 10 $\mu$m on the top surface of the
films [2].  Probably there was contact with only one conducting channel
in these measurements, but in any case there could not have been many
channels involved.  We presume that the reason for the larger critical
currents in the pulsed measurements is that any magnetic flux
associated with the current does not have time to enter the channel,
and so critical currents in this case are determined by other factors,
and may equal depairing currents.

\noindent
{\bf 3. DISCUSSION}

The use of two different values of $(b/a)$ is a simplification of a
model with a quasi continuous distribution of $(b/a)$'s.  With a quasi
continuous distribution, rounding out of corners of the magnetisation
curves would occur.

Since a transition temperature greater than room temperature is very
high for a superconductor, we expect a small coherence length and Type
II superconductivity.   Also, because of the high temperature and
softness of the material, pinning of flux may be difficult.  Thus our
model may be appropriate.  A value of about 5260 Oe is inferred from
the data for the lower critical field $H_{c1}$ which would exist for
fields perpendicular to the superconducting channels if a metamagnetic
transition did not intervene at a lower field .  At 4.2 K, high
conductivity does not disappear for fields up to 9 tesla [21] (probably
approximately parallel to the channel lengths), and so $H_{c2}$ at this
temperature for some orientation of the field is greater than 9 T.
Since $T_c$ has been estimated indirectly to be greater than 700 K in
[2], the low-temperature critical fields may be close to those at room
temperature.  For the high-temperature oxide superconductor
YBa$_2$Cu$_3$O$_7$, $H_{c1}$ for fields perpendicular to the film
planes is about 700 Oe at low $T$ [22].  It would not be surprising to
find much higher values of $H_{c1}$ in oxidised atactic polypropylene
in view of the much higher $T_c$.

We have ignored interaction between current loops, and between current
loops and the larger amount of positively magnetised material in
channels not forming closed loops.  Such interactions can on average be
taken into account by demagnetisation fields.  Since the maximum
susceptibility is always below 1\% of that corresponding to a full
Meissner effect, such corrections can be expected to be represented by
changes in  the internal field with respect to the applied field by
amounts equal to a fraction of a percent of the applied field.

We know of no non-superconducting material which can show diamagnetism
as large as a few tenths of a percent of a full Meissner effect.  For
fields along the $c$-axis, graphite  has a susceptibility of $21.1
\times 10^{-6}$ emu g$^{-1}$ [23], which corresponds to 0.06 \%
of a complete Meissner effect, about an order of magnitude smaller than
the initial susceptibility for the more magnetic of the two films
discussed here.

Although ballistic transport of electrons in arrays of small current
loops may give some magnetic properties similar to those of
superconductors [24], the characteristic temperature $T^{*}$ below which
ballistic effects occur for a loop with $N$ electrons and radius $R$ is
[24] 
\begin{equation} 
k_B T^{*} \sim \hbar^2 N/2m R^2, 
\end{equation} 
with $m$ the electron mass.
Taking $N \sim R/a_0$, where $a_0$ is the period of an assumed periodic
system, this reduces to $k_B T^{*} \sim (\hbar^2/2m)(1/a_0 R)$.
Assuming e.g $R = 1 \mu$m and $a_0=0.5$ nm, we find $T^{*} \sim 0.9$
K.  To move $T^{*}$ up to room temperature, we would thus need to
reduce $R$ to 3 nm, which would imply a radius of the material of the
loop of $\sim$ 1 nm or less.  Although we could obtain an initial
diamagnetic susceptibility as high as 0.7\% of a Meissner effect by
postulating very large numbers of such tiny loops and using Eq.(7),
such small sizes would be incompatible with what is known about
conducting channels through the films, which have cross sections with
radii of the order of 1 $\mu$m [2].  Thus it appears that an
interpretation of the diamagnetism in terms of ballistic effects in
mesoscopic systems at room temperature is very unlikely.

While our model for the diamagnetism associated with
superconducting loops has several arbitrary parameters, and may not be
the only type of superconductor model which can explain the
diamagnetism, we have shown that a plausible model to interpret the
diamagnetic properties can be found based on concepts similar to those
used to explain the strong evidence for superconductivity in channels
from the three types non-magnetic experiments mentioned at the
beginning of the paper.

One way to test the model discussed here would be to cycle the magnetic
field in a suitable sample, keeping within a field range below that at
which the metamagnetic transition occurs, and to compare the
predictions for the complete cycle with Shoenberg's theory [18].
However, since only about 10\% of the samples with conducting channels
are reported to show large diamagnetism [6], finding a suitable film
for such measurements may not be easy.  The only recent reports of work
on oxidised atactic polypropylene come from Professor Grigorov and
members or ex members of his group, several of whom are now working for
commercial companies in the USA, and from Shlimak and Martchenkov [25],
of Bar-Ilan University, who have done more work on
polydimethylsiloxane (PDMS), (which also shows some of the unusual properties
found in OAPP [20,26]), but also some on oxidised atactic
polypropylene.  Besides having some room-temperature properties similar
to those of OAPP, the Josephson effect has been observed for PDMS
sandwiched between two superconductors  at temperatures below $T_c$ of
the superconducting contacts [27].

Another type of polymer which shows narrow channels through films with
fairly high conductivity is poly(3,3'-phthalidylene-4,4-biphenylylene)
(PPB) (see e.g. [28]).  Although no suggestions that these channels are
superconducting at high temperature have been published,
resistance too low to be detected has been found through PPB films with
Sn contacts at low temperatures [29].  Josephson effects have been
reported in polyimide [30].  Further studies of both these
materials, especially of magnetic properties, may be worthwhile.

Two other examples of what may be quasi one-dimensional systems with
superconductivity at room temperature are (i) carbon deposits
containing multiwalled nanotubes [31-33], and (ii) powdered mixtures of
PbCO$_3$.2PbO $+$ Ag$_2$O [34].  The structure of the superconducting
components of the system studied in [34] has been suggested to contain
well separated Ag-O chains which are thought to be the main channels
for possible superconductivity in this system.

\noindent
{\bf 4. CONCLUSIONS}

Fair fits to the magnetisation curves showing large diamagnetism at low
fields at room temperature in two samples of  films of oxidised atactic
polypropylene have been obtained using a model involving
superconducting current loops in applied magnetic fields below those
at which metamagnetic transitions occur, and also below the field
$H_{c1}$ at which resistance would appear assuming no pinning if a
metamagnetic transition did not set in first.  Two sizes of loops with
the same radius $a$ of the cross section of the superconducting
channels of the loops are assumed.  For the second sample, $a$ has been
estimated from the fit to be less than about $0.8 \mu$m,  in fair
agreement with $a < 1 \mu$m estimated from other data [2,17].

\vskip 2em
\noindent
{\bf ACKNOWLEDGMENT}

I wish to thank Professor L.N. Grigorov for
correspondence about the results given in [7].

\noindent
{\bf REFERENCES}

\begin{description}
\item$\;${1.} V.M. Arkhangorodski\u{i}, A.N. Ionov, V.M. Tuchkevich, 
and I.S. Shlimak, {\em Pis'ma Zh. Eksp. Teor. Fiz.} {\bf 51}, 56 (1990)
[{\em JETP Lett.} {\bf 51}, 67 (1990)].
\item$\;${2.} O.V. Demicheva, D.N. Rogachev, S.G. Smirnova, E.I. Shklyarova,
M.Yu. Yablokov, V.M. Andreev, L.N. Grigorov, {\em Pis'ma Zh. Eksp. Teor.
Fiz.} {\bf 51}, 228 (1990) [{\em JETP Lett.} {\bf 51}, 258 (1990)]. 
\item$\;${3.} L.N. Grigorov, O.V. Demicheva, S.G. Smirnova, 
{\em Sverkhprovodimost' (KIAE)} {\bf 4}, 399 (1991) [{\em Superconductivity, Phys.
Chem. Tech.} {\bf 4}, 345 (1991)].
\item$\;${4.} A.V. Krayev, T.V. Dorofeeva, E.I. Shklyarova, L.N.
Grigorov, 9th. CIMTEC - World Forum on New Materials, Florence, Italy,
14-19 June 1998; {\em Advances in Science and Engineering Technology, Vol.
23: Science and Engineering of HTC Superconductivity, edited by P. Vincennzini}
(Faenza Techna., Srl., 1999), pp. 459-466.
\item$\;${5.} S.G. Smirnova, O.V. Demicheva, L.N. Grigorov,
{\em Pis'ma Zh. Eksp. Teor. Fiz.} {\bf 48}, 212 (1988) [{\em JETP Lett.}
{\bf 48}, 231 (1988)].
\item$\;${6.} N.S. Enikolopyan, L.N. Grigorov, S.G. Smirnova,
{\em Pis'ma Zh. Eksp. Teor. Fiz.} {\bf 49}, 326 (1989) [{\em JETP Lett.}
{\bf 49}, 371 (1989)]. 
\item$\;${7.} D.N. Rogachev, L.N. Grigorov, {\em J. Supercond.} {\bf
13}, 947 (2000).
The straight dashed lines through the origin shown in Fig. 1 of this paper
have smaller slopes than indicated by the labelling for $\chi$.  The values
for $\chi$ given are based on an assumed form of extrapolation of the 
observed moments to lower fields than those to which they were
measured (L.N. Grigorov, personal communication).
\item$\;${8.} L.N. Grigorov, D.N. Rogachev, A.V. Kraev, {\em Vysokomol.
Soedin. B} {\bf 35}, 1921 (1993) [{\em Polymer Science} {\bf 35}, 1625 (1993)].
\item$\;${9.} A.M. Elyashevich, A.A. Kiselev, A.V. Liapzev,
G.P. Miroshnichenko, {\em Phys. Lett. A} {\bf 156}, 111 (1991).
\item{10.} D.M. Eagles, {\em Physica C} {\bf 225}, 222 (1994); erratum 
{\em ibid.} {\bf 280}, 335 (1997).
\item{11.} D.M. Eagles, {\em J. Supercond.} {\bf 11}, 189 (1998).  On p. 192 of
this paper, the magnetic moments of two samples inferred from published
data were too large by a factor of $4\pi$, besides having some
uncertainty because of lack of precise knowledge by me at that time of
the volumes of the samples.  This implies that the radii of loops
required to fit the moments on the assumption of spontaneous currents
close to the critical current can be smaller by a factor of the order
of $(1/4\pi)^{\frac{1}{2}}$ than given in the reference.  In fact,
judging by the magnetisations given in the recent paper for a sample
previously reported on in [6], the moments estimated in [11] for that
sample are too large by another factor of two, which appears to have
arisen because the low-field susceptibility given in [6] was, I
infer from correspondence from Professor Grigorov about reference
[7], based partly on an assumed extrapolation of the measured moment
curves to lower fields than those for which measurements were made.  The
smaller moments required from those estimated in [11] give rise to the
possibility of alternative explanations in terms of smaller induced
currents as discussed in the present paper for the results of [7].
\item{12.} D.M. Eagles, {\em Revista Mexicana de Fisica} {\bf 45}, {\em
Suplemento 1}, 118-121 (1999).
\item{13.} L.N. Grigorov, {\em Phil. Mag. B} {\bf 78}, 353 (1998).
\item{14.} L.N. Grigorov,  9th. CIMTEC - World Forum on
New Materials, Florence, Italy, 14-19 June 1998; {\em Advances in Science
and Engineering Technology, Vol. 23: Science and Engineering of HTC
Superconductivity, edited by P.Vincennzini} (Faenza Techna. Srl., 1999), pp.
675-684.
\item{15.} L.N. Grigorov, D.N. Rogachev,  {\em Molec. Cryst. Liquid Cryst.}
{\bf 230}, 625 (1993). 
\item{16.} S.G. Smirnova, E.I. Shklyarov, L.N. Grigorov,
{\em Vyosokomol. Soedin. B} {\bf 31}, 667 (1989). 
\item{17.} V.M. Arkhangorodski\u{i}, E.G. Guk, A.M. El'yashevich, 
A.N. Ionov, V.M. Tuchkevich, I.S. Shlimak,  {\em Dokl. Akad. Nauk. SSSR}
{\bf 309}, 603 (1989) [{\em Sov. Phys. Doklady} {\bf 34}, 1016 (1989)].
\item{18.} D. Shoenberg, {\em Superconductivity} (Cambridge University
Press, 1952) Sec.2.6.  Note that in Fig. 12 of this reference, the
arbitrary units of magnetic moment used appear to be negative, i.e.
opposed to the applied field, as expected.  This can be seen from
Shoenberg's equations.
\item{19.} W.H. Press, B.P. Flannery, S.A. Teukolsky, W.T. Vetterling,
{\em Numerical Recipes} (Cambridge University Press, 1986).
\item{20.} A.V. Kraev, S.G. Smirnova, L.N. Grigorov, {\em Vysokomol. Soedin.
A} {\bf 35}, 1308 (1993). [{\em Polymer Science} {\bf 35}, 1308 (1993)].
\item{21.} I.S. Shlimak, L.N. Grigorov, unpublished results, 1996,
referred to in [7].
\item{22.} S. Senoussi, C. Aguillon, {\em Europhys. Lett.} {\bf 12}, 273 (1990).
\item{23.} X.K. Wang, R.P.H. Chang, A. Patachinski, J.B. Ketterson,
{\em J. Mater. Res.} {\bf 9}, 1578 (1994).
\item{24.} M. Szopper, E. Zipper, {\em Int. J. Mod. Phys. B} {\bf 9},
161 (1995).
\item{25.} I. Shlimak, V. Martchenkov, {\em Solid State Commun.} {\bf 107},
443 (1998).
\item{26.} L.N. Grigorov, T.V. Dorofeeva, A.V. Kraev, D.N. Rogachev,
O.V. Demicheva, E.I. Shklyarova, {\em Vysokomol. Soedin A} {\bf 38},
(1996) 2011.  [{\em Polymer Science A} {\bf 38}, (1996) 1328].
\item{27.} A.N. Ionov, V.A. Zakrevski\u{i}, {\em Pis'ma Zh. Tekh.
Fiz.} {\bf 26}, No. 20, p. 36 (2000).  [{\em Tech. Phys. Lett.} {\bf 26}, 910 
(2000)].
\item{28.} V.M. Kornilov, A.N. Lachinov, {\em Synth. Met.} {\bf 53},
71 (1992).
\item{29.} V.A. Zakrevski\u{i}, A.N. Ionov, A.N. Lachinov,
{\em Pis'ma Zh. Tekh. Fiz.} {\bf 24}, No. 13, p. 89 (1998). [{\em Tech.
Phys. Lett.} {\bf 24}, 539 (1998)].
\item{30.} A.N. Ionov, V.A. Zakrevski\u{i}, I.M. Lazebnik, {\em Pis'ma
Zh. Tekh. Fiz.} {\bf 25}, No. 17, p. 36 (1999). [{\em Tech. Phys. Lett.} 
{\bf 25}, 691 (1999)].
\item{31.} V.I. Tsebro, O.E. Omel'yanovski\u{i}, A.P. Moravski\u{i},
{\em Pis'ma Zh. Eksp. Teor. Fiz.} {\bf 70}, 457 (1999). [{\em JETP
Lett.} {\bf 70}, 462 (1999)].
\item{32.} V.I. Tsebro, O.E. Omel'yanovski\u{i}, {\em Phys. Usp.}
{\bf 43}, 847 (2001).
\item{33.} G.-M. Zhao, Y.S. Wang, {\em cond-mat 011268} (2001); {\em
Phil. Mag. B,} to be published.
\item{34.} D. Djurek, Z. Meduni\'{c}, A. Tonejc, M. Paljevi\'{c},
{\em Physica C} {\bf 351}, 78 (2001).

\end{description}

\noindent
{\bf Figure Caption}

\noindent
{\bf Fig. 1.}  Comparison of model calculations with diamagnetism associated
with closed loops inferred from experiment for applied fields $H_e$
less than the fields at which the metamagnetic transition occurs, i.e.
$H_e <$ 3360 Oe for sample 1, and $H_e <$ 1390 Oe for sample 2.  The
diamagnetic contribution to the net magnetisation has been inferred
from the observed net magnetisation by correction for a probable
positive contribution from conducting channels which do not form closed
loops (see text).  
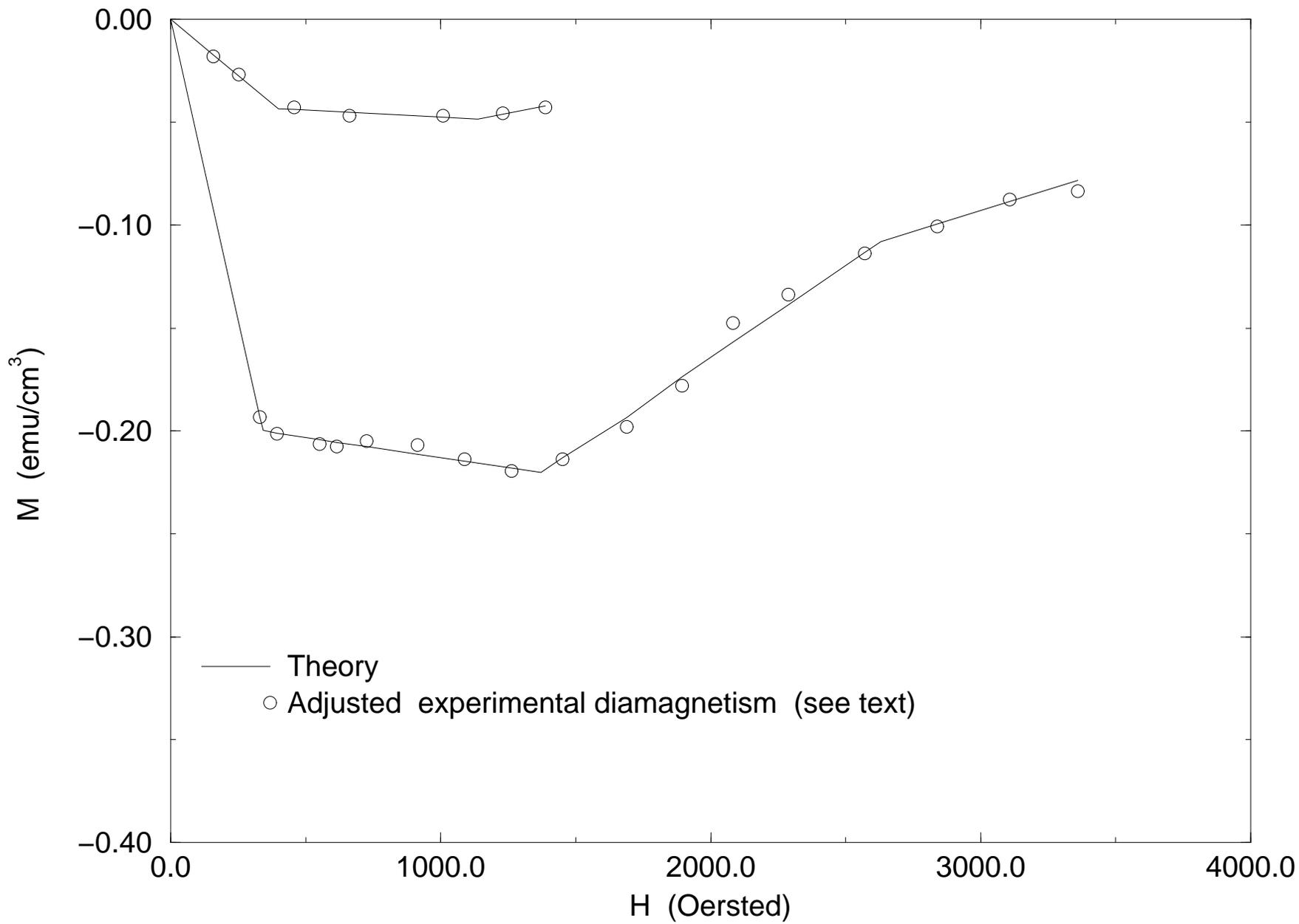
\end{document}